%% file: neurips_2020.tex
\documentclass{article}
\pdfoutput=1

\usepackage[nonatbib,final]{neurips_2020_ml4ps}

\usepackage[utf8]{inputenc} 
\usepackage[T1]{fontenc}    
\usepackage{hyperref}       
\usepackage{url}            
\usepackage{booktabs}       
\usepackage{amsfonts}       
\usepackage{nicefrac}       
\usepackage{microtype}      

\usepackage{amsmath}
\usepackage{bm}
\usepackage{pgfplots}
\pgfplotsset{width=7.0cm, compat=1.15}
\usepgfplotslibrary{groupplots}

\title{Normalizing Flows as a Novel PDF Turbulence Model}

\author{%
  Deniz A.~Bezgin\thanks{Corresponding author.} \\
  Chair of Aerodynamics and Fluid Mechanics\\
  Technical University of Munich\\
  Garching bei M\"unchen \\
  \texttt{deniz.bezgin@tum.de} \\
  \And
  Nikolaus A.~Adams \\
  Chair of Aerodynamics and Fluid Mechanics \\
  Technical University of Munich\\
  Garching bei M\"unchen \\
  \texttt{nikolaus.adams@tum.de} \\
}

\begin{document}

\maketitle

\begin{abstract}
In this paper, we propose normalizing flows (NF) as a novel probability density function (PDF) turbulence model (NF-PDF model) for the Reynolds-averaged Navier-Stokes (RANS) equations.
We propose to use normalizing flows in two different ways: firstly, as a direct model for the Reynolds stress tensor,
and secondly as a second-moment closure model, i.e. for modeling the pressure-strain and dissipation tensor in the Reynolds stress transport equation. 
In classical PDF closure models, a stochastic differential equation has to be modeled and solved to obtain samples of turbulent quantities.
The NF-PDF closure model allows for direct sampling from the underlying probability density functions of fluctuating turbulent quantities,
such that ensemble-averaged quantities can then be computed.  
To illustrate this approach we demonstrate an application for the canonical case of homogeneous shear turbulence.
Training data are extracted from high-fidelity direct numerical simulations (DNS).
\end{abstract}

\section{Introduction}
\label{sec:intro}
Turbulent flows are omnipresent in nature and engineering applications. 
Modeling and simulating turbulent flows have been a major challenge for decades. 
Complex spatio-temporal dynamics and strongly nonlinear behavior over multiple scales make direct numerical simulations
of turbulent flows prohibitively expensive. 
Therefore, the development of reliable turbulence models with high predictive power is as important as ever before.

The Reynolds-averaged Navier-Stokes (RANS) equations are widely used in practical applications.
Closure of the RANS equations is traditionally achieved by employing the turbulent viscosity hypothesis and isotropic closures, such as the prominent $k-\epsilon$ model. 
As such models generally fail to model anisotropy, the more fundamental Reynolds stress transport equations, so called second-moment closures, are used to provide higher fidelity \cite{Pope2000}.
Probability density function (PDF) methods provide an alternative modeling approach in which a transport equation
for the one-point, one-time PDF of turbulent fluctuating quantities is solved \cite{Pope1985, Jenny2001b}.

In recent years, with the emergence of machine learning, promising data-driven turbulence models have been developed \cite{Duraisamy2019}. 
Duraisamy and coworkers \cite{Tracey2013} have proposed deep learning enhancements of established RANS closure models, 
while Ling et al. \cite{Ling2016} put forward a Galilean invariant neural network architecture for modeling the Reynolds stress anisotropy tensor.
Beck et al. \cite{Beck2019} used Deep Neural Networks for LES turbulence modelling, 
while Novati \& Koumoutsakos \cite{Novati2020} used reinforcement learning to adjust the Smagorinsky constant in LES simulations.

In this work, we propose a novel data-driven approach to RANS turbulence modelling which is in spirit close to the PDF modeling approach.
We apply normalizing flows as a PDF-turbulence closure model (NF-PDF model).
We examine the capabilities of normalizing flows to model the Reynolds stress tensor directly by learning the probability distribution of the fluctuating velocities as well as using 
normalizing flows for the closure of the Reynolds stress transport equation (RSTE).
In the latter approach, the NF-PDF model learns the PDFs of the fluctuating pressure and the spatial gradient of the fluctuating velocities, thereby modeling the pressure-strain correlation and the dissipation term.
We show proof of concept by the example of homogeneous shear turbulence (HST).
Ground truth data are generated by direct numerical simulations.
To the best of our knowledge, this work is the first time normalizing flows are used as PDF closure models. 

In the next section we present the NF-PDF closure model.
Section \ref{sec:HST} presents the governing equations for HST.
In section \ref{sec:results}, we present results of the NF-PDF closure model for bounded and unbounded homogeneous shear turbulence. 
Finally, conclusions are given in section \ref{sec:conclusion}.

\section{Normalizing Flows as a PDF Closure}
\label{sec:method}

\subsection*{Problem Statement}
\label{sec:statement}
The Reynolds-averaged Navier-Stokes equations \eqref{eq:RANS} contain the unclosed Reynolds stresses.
In RANS turbulence modeling, often the Reynolds stresses $R_{ij} = <u_i'u_j'>$ are modeled invoking an eddy viscosity hypothesis, e.g. with a two-equation model.

\begin{align}
  \frac{\partial <u_i>}{\partial t} + \frac{\partial <u_i u_j>}{\partial x_j} = - \frac{\partial <p>}{\partial x_i} + \nu \frac{\partial^2 <u_i>}{\partial x_j^2} - \frac{\partial <u_i'u_j'>}{\partial x_j}
  \label{eq:RANS}
\end{align}

Here, $u$ is the velocity, $p$ is the pressure, and $\nu$ is the kinematic viscosity.
$< (\cdot) >$ represents an ensemble average, while $(\cdot)'$ is a fluctuating quantity.
In Reynolds stress turbulence closure, transport equations for each component of the Reynolds stress tensor are solved.
The unclosed terms in Eq. \eqref{eq:RSTE}, i.e. pressure-strain $\Pi_{ij} = <p' \left( \frac{\partial u_i'}{\partial x_j} + \frac{\partial u_j'}{\partial x_i} \right)>$ and dissipation $\epsilon_{ij} = 2 \nu <\frac{\partial u_i'}{\partial x_k} \frac{\partial u_j'}{\partial x_k} >$, have to be closed. 
$P_{ij}$ is the turbulent production.

\begin{align}
  \frac{\partial R_{ij}}{\partial t} = P_{ij} + \Pi_{ij} - \epsilon_{ij}
  \label{eq:RSTE}
\end{align}

Classical PDF methods model the transport equation for the one-point, one-time PDF of certain fluid properties, e.g. velocity \cite{Pope1985,Haworth1986,Jenny2001b}.
Instead of solving the high-dimensional Fokker-Planck equation directly, Langevin methods are used.
A set of particles is advected by a corresponding stochastic differential equation such that the particles have the same PDF as the underlying turbulent flow.
Averaging over the particle ensembles yields estimates for the right-hand side of Eq. \eqref{eq:RSTE} in closed form given the sample means.

\subsection*{Normalizing Flows}
\label{sec:NF}

In this work, we propose the normalizing flow PDF (NF-PDF) closure model for the RANS equations.
The key idea of the NF-PDF model is to learn the underlying probability distributions of turbulent flow properties.
Once the PDF is learned, straightforward sampling and successive ensemble-averaging from the NF model allows for very fast evaluation
of the unclosed turbulent terms.
Our approach circumvents the solution of stochastic differential equations (as in classical PDF methods) and can incorporate anisotropic flow conditions.
The model also avoids adhoc closures that are part of state-of-the-art Reynolds stress models. 
A conditional NF-PDF model allows for time-dependent PDF, or PDF which are subject to given mean flow conditions.
The NF-PDF model extends to high-dimensional problems enabling also to learn complex distributions over multiple variables,
e.g. chemical compositions in turbulent reactive flows.

Normalizing flows are able to learn almost any probability distribution from data \cite{Dinh2014,Rezende2015,Dinh2017}.
By applying a series of bijective transformations $g=f^{-1}$ to a simple prior probability distribution $p_Z(\bm{z})$,
increasingly complex target distributions $p_X(\bm{x})$ can be approximated.
Given the change of variable formula,

\begin{align}
  \log (p_X(\bm{x} \vert \bm{c}, \theta )) = \log (p_Z(f(\bm{x} \vert \bm{c}, \theta ))) + \log \left| \det \left(\frac{\partial f(\bm{x} \vert \bm{c}, \theta )}{\partial \bm{x}^T} \right) \right|,
  \label{eq:changevariables}
\end{align}

the model can be trained by minimizing the negative log-likelihood.
In Eq. \eqref{eq:changevariables}, $\theta$ are the trainable parameters of the flow, $\bm{c}$ is a conditioning argument, and $\frac{\partial f(\bm{x} \vert \bm{c}, \theta )}{\partial \bm{x}^T}$ represents the Jacobian of $f$. 
At sampling time, a sample is drawn from the prior probability distribution $p_Z$.
Applying the inverse transformation $f^{-1}$ generates a sample in the original space $\bm{x} \sim p_X$ .

Normalizing flows are often parameterized by neural networks which allow highly flexible transformations.
In this work, we parameterize our normalizing flow with a sequence of real-valued non-volume preserving (real NVP) transformations \cite{Dinh2017}. 
RNVP makes use of affine coupling layers. 
The input to a RNVP layer $\bm{z} \in \mathbb{R}^D$ is split into two disjoint parts $(\bm{z}^A, \bm{z}^B) \in \mathbb{R}^d \times \mathbb{R}^{D-d}$.
Then, the output of a single RNVP layer is calculated as 

\begin{align}
  \bm{x}^A &= \bm{z}^A \\
  \bm{x}^B &= \bm{z}^B \odot \exp\left( s(\bm{z}^A) \right) + t\left( \bm{z}^A \right),
  \label{eq:affinecoupling}
\end{align}

where $s$ and $t$ are neural networks and $\odot$ denotes element-wise multiplication.
The Jacobian of said transformation is triangular which allows for efficient calculation of its determinant.
We use an isotropic Gaussian as the prior $p_Z(\bm{z})$.
A conditional version of the affine coupling layer was established in \cite{Ardizzone2019} which allows for conditional sample creation.
The condition $\bm{c}$, e.g. mean flow conditions or time, is simply concatenated with the relevant input to the neural networks $\bm{z}^A$ in Eq. \eqref{eq:affinecoupling}.

\section{Homogeneous Shear Turbulence (HST)}
\label{sec:HST}
The reference dataset is generated from a $3D$ DNS of HST.
HST is a canonical turbulent flow. 
The presence of a mean velocity gradient without disturbances from boundary effects makes HST the simplest non trivial turbulent shear flow.
Experimental studies have been conducted by Corrsin and coworkers \cite{Champagne1970, Tavoularis1981}.
Since Rogallo's pioneering work \cite{Rogallo1981} many numerical investigations followed \cite{Rogers1987,Pumir1996,Isaza2009}.
We employ Rogallo's method \cite{Rogallo1981} and write the incompressible Navier-Stokes equations in a frame of reference
moving with the mean shear flow $<u_1> = S x_2$, 

\begin{align}
  \frac{\partial u_i'}{\partial x_i^*} - St \frac{\partial u_2'}{\partial x_1^*} &= 0, \\
  \frac{\partial u_i'}{\partial t} + S \delta_{i1} u_2 + u_j' \frac{\partial u_i'}{\partial x_j^*} - St u_2' \frac{\partial u_i'}{\partial x_1^*} &= - \frac{\partial p'}{\partial x_i^*} + \nu \left( \frac{\partial^2 u_i'}{\partial (x_k^*)^2} + S^2t^2 \frac{\partial^2 u_i'}{\partial (x_1^*)^2} - 2St \frac{\partial^2 u_i'}{\partial x_1^* \partial x_2^*} \right),
  \label{eq:nse}
\end{align}

where $u_i'$ is the fluctuating velocity, $p'$ is the fluctuating pressure, and $S$ is the shear rate.
The comoving reference frame is related to the stationary reference frame by $x_i^{*} = x_i - St \delta_{i1} x_2.$
Rogallo's transformation removes the explicit dependence of the Navier-Stokes equations on the spatial coordinate and 
allows the imposition of periodic boundary conditions.
The equations are solved with an in-house parallelized pseudo-spectral code.

\section{Results}
\label{sec:results}

\subsection{Direct Reynolds Stress Tensor Modeling}
\label{sec:BHST}
\begin{figure}
  \centering
  \input{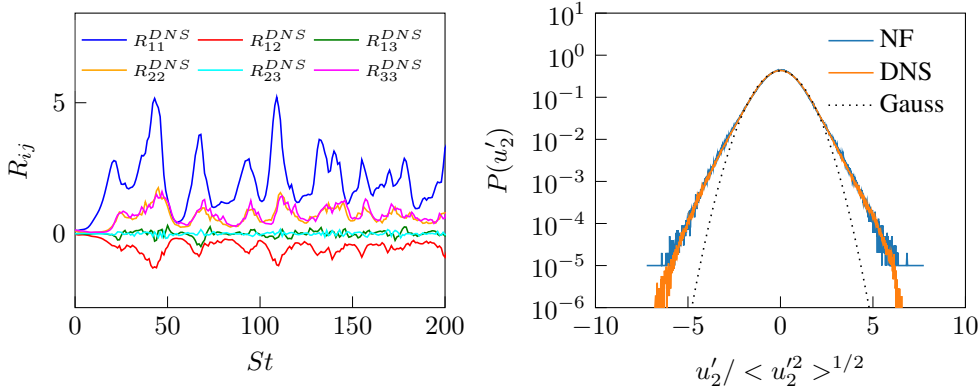}
  \caption{Left: Reynolds stress tensor over normalized time. Right: PDF of the $u_2'$-velocity component with $n_{sample}=10^{7}$.
  The dotted line is a Gaussian distribution with same mean and variance as the DNS data. Parameters of the HST flow are: $S=1, \nu=0.0015, Re=SL^2/\nu = 2632$.}
  \label{fig:BHST}
\end{figure}

The integral length scale in HST grows over time.
Once it reaches the scale of the numerical domain, a statistically steady state can be reached in which turbulent energy production by mean shear 
and viscous energy dissipation are balanced.
Fig. \ref{fig:BHST} shows the Reynolds stress components which become statistically stationary after an intial transient.
Statistics are gathered once a statistically stationary state is reached.

We train an unconditional normalizing flow to learn the probability density distribution of the fluctuating velocities $u_i'$, i.e. $\bm{x} = \bm{u}' \in \mathbb{R}^{3}$.
The fluctuating velocities show non Gaussian behavior in the $x_2$ and $x_3$ component, see for example Fig. \ref{fig:BHST} (right).
Sampling from the NF model gives us $u_i'$, so that $R_{ij} = <u_i'u_j'>$ can be easily computed.
A RNVP flow of $8$ coupling blocks is used.
The neural networks in the affine coupling layers have three hidden layers with $64$ neurons. 

The NF accurately captures the PDF of the velocity components.
By sampling from the NF and successive ensemble-averaging the Reynolds stress components can be calculated.
Table \ref{tab:bij} shows very accurate predictions for the Reynolds anisotropy tensor $b_{ij} = <u_i'u_j'>/<u_k'u_k'> - \delta_{ij} /3$ 
when compared to the DNS results.

\begin{table}
  \caption{Components of the velocity anisotropy tensor $b_{ij}$.}
  \label{tab:bij}
  \centering
  {\renewcommand{\arraystretch}{1.1}
  \begin{tabular}{c c c c c}
    \hline \hline 
    Case & $b_{11}$ & $b_{22}$ & $b_{33}$ & $b_{12}$ \\ 
    \hline 
    DNS & $0.245$ & $-0.128$ & $-0.118$ & $-0.159$ \\
    NF, $n_{sample} = 10^3$ & $0.246$ & $-0.126$ & $-0.119$ & $-0.166$ \\
    NF, $n_{sample} = 10^5$ & $0.245$ & $-0.125$ & $-0.120$ & $-0.160$ \\
    \hline
  \end{tabular}
  }
\end{table}

\subsubsection{Closure of the Reynolds Stress Transport Equation}
\label{sec:UHST}

In the initial transient state of HST, the Reynolds stresses are time dependent.
We use a conditional NF to estimate the pressure-strain and dissipation tensor, 
which can then be used in the RSTE \eqref{eq:RSTE} to compute the Reynolds stresses.
We learn the PDF of the velocity derivatives $p(\frac{\partial u_i'}{\partial x_j}\vert t)$ and of the fluctuating pressure $p(p'\vert t)$ conditioned on time $t$,
i.e. $\bm{x} = \left(\frac{\partial u_i'}{\partial x_j}, p' \right) \in \mathbb{R}^{10}$.
An $8$ layers RNVP flow is constructed with neural networks with three hidden layers each containing $128$ hidden units.
The time $t$ is concatenated to $\bm{x}$ and passed as an input to each RNVP layer.\\

\begin{figure}[h]
  \centering
  \input{NF_eps_pi_100000.tex}
  \caption{Pressure-strain correlation $\Pi_{ij}$ (left) and dissipation $\epsilon_{ij}$ (right) over normalized time.
  The initialized velocity field is incompressible and isotropic. 
  The NF-PDF model is fitted once a natural turbulent velocity field has developed ($St=2$).
  $S=20\sqrt{2}, \nu=0.01/\sqrt{2}, Re=SL^2/\nu = 157914$.}
  \label{fig:UHST}
\end{figure}
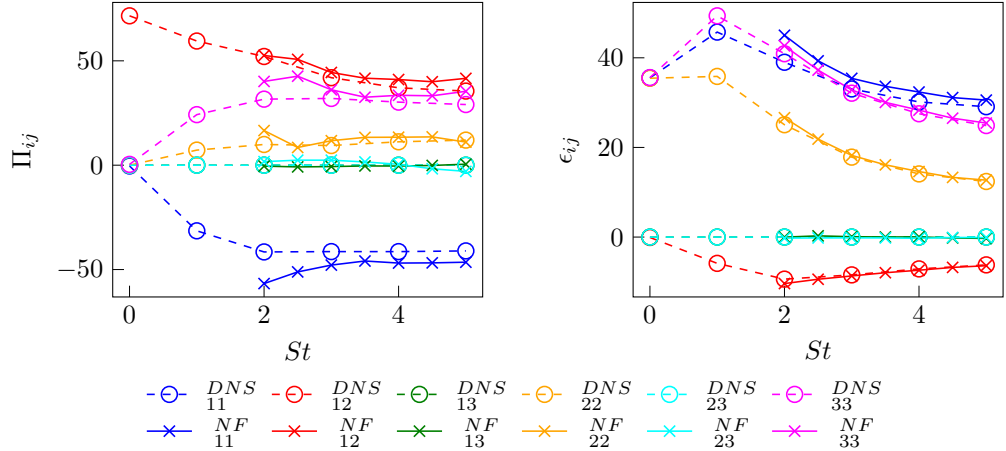

The NF-PDF model accurately captures the dissipation term while the pressure-strain tensor prediction is adequate, see Fig. \ref{fig:UHST}. 
In addition, the NF-PDF model yields very good predictive results when evaluated at time instances in between training points from the DNS. 
We observe a slight deviation from the DNS results in the $\Pi_{11}$ component at early times.
This error decreases with time.  

\section{Conclusion}
\label{sec:conclusion}

We have presented a novel turbulence closure model on the basis of normalizing flows.
In this proof of concept, we have shown that the NF-PDF closure model can successfully model the Reynolds stress tensor as well as
the pressure-strain correlation and the dissipation tensor.
As the pressure-strain tensor represents non-local interactions, its accurate prediction is particularly challenging.
NF-PDF models have the potential to solve this key challenge of turbulence.
Further research focuses on improving the prediction fidelity and on extending our method to more complex turbulent flows.

\section*{Broader Impact}
Turbulent flows are omnipresent in nature and engineering applications.
On the one hand, undestanding turbulent flows plays an important role in the design of efficient engineering devices.
For example, wind turbines rely heavily on the profound understanding of the incoming flow conditions, 
while turbulent mixing reduces emissions of combustion engines. 
On the other hand, the accurate simulation of turbulent atmospheric flows is crucial for weather forecasting.
High-fidelity turbulence models play a central role in the prediction of hazardous weather events, e.g. hurricanes, air pollution, and wild fires.
The authors recognize that their work contributes to both aforementioned aspects.
Although the current work is still very much fundamental turbulence research, the authors are aware of its societal impact.
At this point, the authors do not see any ethical concerns regarding the present work.

%


\end{document}

%% file: NF_eps_pi_100000.tex
\begin{tikzpicture}

\definecolor{color0}{rgb}{1,0.647058823529412,0}
\definecolor{color1}{rgb}{0,1,1}
\definecolor{color2}{rgb}{1,0,1}

\begin{groupplot}[group style={group size=2 by 1, horizontal sep=2.0cm}]
\nextgroupplot[
width=6.5cm, height=5.5cm,
tick align=inside,
tick pos=left,
x grid style={white!69.0196078431373!black},
xlabel={\(\displaystyle St\)},
xmin=-0.250000000000007, xmax=5.25000000000014,
xtick style={color=black},
y grid style={white!69.0196078431373!black},
ylabel near ticks,
ylabel={\(\displaystyle \Pi_{ij}\)},
ymin=-63.1338896489193, ymax=77.9807723781159,
ytick style={color=black}
]
\addplot [semithick, blue, dashed, mark=o, mark size=3, mark options={solid}]
table {%
0 -0.375908214883965
1 -31.4475143983725
1.99999999999999 -41.5628756697825
3.00000000000004 -41.4031430844778
4.00000000000009 -41.4104339555493
5.00000000000014 -41.0706327839499
};
\addplot [semithick, red, dashed, mark=o, mark size=3, mark options={solid}]
table {%
0 71.5664695587052
1 59.4899675328203
1.99999999999999 52.0775997538661
3.00000000000004 41.9233459399942
4.00000000000009 37.0789130378089
5.00000000000014 35.5072420929077
};
\addplot [semithick, green!50.1960784313725!black, dashed, mark=o, mark size=3, mark options={solid}]
table {%
0 0
1 0
1.99999999999999 0
3.00000000000004 0
4.00000000000009 0
5.00000000000014 0
};
\addplot [semithick, color0, dashed, mark=o, mark size=3, mark options={solid}]
table {%
0 -0.0128286417588088
1 7.23068727562633
1.99999999999999 9.92225272261003
3.00000000000004 9.42139357851762
4.00000000000009 11.1996018098309
5.00000000000014 12.0128114147392
};
\addplot [semithick, color1, dashed, mark=o, mark size=3, mark options={solid}]
table {%
0 0
1 0
1.99999999999999 0
3.00000000000004 0
4.00000000000009 0
5.00000000000014 0
};
\addplot [semithick, color2, dashed, mark=o, mark size=3, mark options={solid}]
table {%
0 0.388736856642774
1 24.2168271227632
1.99999999999999 31.6406229471869
3.00000000000004 31.9817495059685
4.00000000000009 30.210832145726
5.00000000000014 29.0578213692176
};
\addplot [semithick, blue, mark=x, mark size=3, mark options={solid}]
table {%
1.99999999999999 -56.7195868295086
2.50000000000001 -51.1340549597168
3.00000000000004 -47.862007477583
3.50000000000006 -45.9100393676527
4.00000000000009 -46.8925726600221
4.50000000000011 -46.8033704551157
5.00000000000014 -46.478939530265
};
\addplot [semithick, red, mark=x, mark size=3, mark options={solid}]
table {%
1.99999999999999 52.5636864915675
2.50000000000001 50.6973091447849
3.00000000000004 44.4201439691267
3.50000000000006 41.6740858259488
4.00000000000009 41.0483856742443
4.50000000000011 39.9760500411285
5.00000000000014 41.6068759707015
};
\addplot [semithick, green!50.1960784313725!black, mark=x, mark size=3, mark options={solid}]
table {%
1.99999999999999 -0.451267249718525
2.50000000000001 -0.821699944909791
3.00000000000004 -0.666245908073738
3.50000000000006 -0.360704700326001
4.00000000000009 -0.410517739263197
4.50000000000011 -0.123182891350072
5.00000000000014 0.481484523161322
};
\addplot [semithick, color0, mark=x, mark size=3, mark options={solid}]
table {%
1.99999999999999 16.5906799880737
2.50000000000001 8.4926990170123
3.00000000000004 11.7286417847122
3.50000000000006 13.3386322627253
4.00000000000009 13.4070981021133
4.50000000000011 13.5589926226542
5.00000000000014 11.0715802546363
};
\addplot [semithick, color1, mark=x, mark size=3, mark options={solid}]
table {%
1.99999999999999 1.70372004143899
2.50000000000001 2.40680917098789
3.00000000000004 2.39791917882683
3.50000000000006 1.55329228787501
4.00000000000009 0.4268268431482
4.50000000000011 -1.72510979336373
5.00000000000014 -2.97186076022935
};
\addplot [semithick, color2, mark=x, mark size=3, mark options={solid}]
table {%
1.99999999999999 40.129123450716
2.50000000000001 42.6414452437706
3.00000000000004 36.1333207569151
3.50000000000006 32.5715370925181
4.00000000000009 33.4855442194925
4.50000000000011 33.244414201498
5.00000000000014 35.4074032426738
};

\nextgroupplot[
width=6.5cm, height=5.5cm,
tick align=inside,
tick pos=left,
x grid style={white!69.0196078431373!black},
xlabel={\(\displaystyle St\)},
xmin=-0.250000000000007, xmax=5.25000000000014,
xtick style={color=black},
y grid style={white!69.0196078431373!black},
ylabel={\(\displaystyle \epsilon_{ij}\)},
ymin=-13.3626780904732, ymax=52.3538015810056,
ytick style={color=black},
legend style = { column sep = 3pt, legend columns = 6, legend to name = grouplegend, draw=none}]
]
\addplot [semithick, blue, dashed, mark=o, mark size=3, mark options={solid}]
table {%
0 35.5277477836016
1 45.7394647086181
1.99999999999999 39.0083358356276
3.00000000000004 33.0496503563117
4.00000000000009 30.1499023684193
5.00000000000014 29.0891389550125
};
\addlegendentry{$_{11}^{DNS}$}
\addplot [semithick, red, dashed, mark=o, mark size=3, mark options={solid}]
table {%
0 -0.0896902287260478
1 -5.86687494338189
1.99999999999999 -9.38713119674603
3.00000000000004 -8.43969539232687
4.00000000000009 -7.10684434046243
5.00000000000014 -6.20723438020361
};
\addlegendentry{$_{12}^{DNS}$}
\addplot [semithick, green!50.1960784313725!black, dashed, mark=o, mark size=3, mark options={solid}]
table {%
0 0
1 0
1.99999999999999 0
3.00000000000004 0
4.00000000000009 0
5.00000000000014 0
};
\addlegendentry{$_{13}^{DNS}$}
\addplot [semithick, color0, dashed, mark=o, mark size=3, mark options={solid}]
table {%
0 35.441695635215
1 35.8469748586963
1.99999999999999 25.0653408146065
3.00000000000004 17.8355850694088
4.00000000000009 14.0897210501067
5.00000000000014 12.4018760603271
};
\addlegendentry{$_{22}^{DNS}$}
\addplot [semithick, color1, dashed, mark=o, mark size=3, mark options={solid}]
table {%
0 0
1 0
1.99999999999999 0
3.00000000000004 0
4.00000000000009 0
5.00000000000014 0
};
\addlegendentry{$_{23}^{DNS}$}
\addplot [semithick, color2, dashed, mark=o, mark size=3, mark options={solid}]
table {%
0 35.5823935434587
1 49.3666888686656
1.99999999999999 40.9069654269576
3.00000000000004 32.1039599584945
4.00000000000009 27.5116710475821
5.00000000000014 24.8835622991451
};
\addlegendentry{$_{33}^{DNS}$}
\addplot [semithick, blue, mark=x, mark size=3, mark options={solid}]
table {%
1.99999999999999 45.027482128176
2.50000000000001 39.3179988388213
3.00000000000004 35.415646648966
3.50000000000006 33.6341520342667
4.00000000000009 32.3649220509561
4.50000000000011 31.1215366694336
5.00000000000014 30.5614339670298
};
\addlegendentry{$_{11}^{NF}$}
\addplot [semithick, red, mark=x, mark size=3, mark options={solid}]
table {%
1.99999999999999 -10.3755653781332
2.50000000000001 -9.41044635978511
3.00000000000004 -8.67676557643434
3.50000000000006 -7.95553884635957
4.00000000000009 -7.39440651264509
4.50000000000011 -6.76838294875806
5.00000000000014 -6.3967275886023
};
\addlegendentry{$_{12}^{NF}$}
\addplot [semithick, green!50.1960784313725!black, mark=x, mark size=3, mark options={solid}]
table {%
1.99999999999999 -0.0142696527779213
2.50000000000001 0.268598737213534
3.00000000000004 0.0840633458099783
3.50000000000006 0.0134911498848267
4.00000000000009 0.0446061555143993
4.50000000000011 -0.20388840766347
5.00000000000014 -0.2996253460347
};
\addlegendentry{$_{13}^{NF}$}
\addplot [semithick, color0, mark=x, mark size=3, mark options={solid}]
table {%
1.99999999999999 26.6250473371689
2.50000000000001 21.8602128859794
3.00000000000004 18.2228123578668
3.50000000000006 16.1147769018061
4.00000000000009 14.6145443502442
4.50000000000011 13.310953815196
5.00000000000014 12.7065496888937
};
\addlegendentry{$_{22}^{NF}$}
\addplot [semithick, color1, mark=x, mark size=3, mark options={solid}]
table {%
1.99999999999999 -0.236438610435987
2.50000000000001 -0.196840703010673
3.00000000000004 -0.178852469551064
3.50000000000006 -0.192250939016891
4.00000000000009 -0.276873457178776
4.50000000000011 -0.0910481624128167
5.00000000000014 -0.10334858777077
};
\addlegendentry{$_{23}^{NF}$}
\addplot [semithick, color2, mark=x, mark size=3, mark options={solid}]
table {%
1.99999999999999 42.6675581416218
2.50000000000001 37.2561573695871
3.00000000000004 32.8300447887325
3.50000000000006 30.0518195859645
4.00000000000009 28.3222044420491
4.50000000000011 26.5274129964675
5.00000000000014 25.4322126069185
};
\addlegendentry{$_{33}^{NF}$}
\end{groupplot}
\node at ($(group c2r1) + (-4.0cm,-3.5cm)$) {\ref*{grouplegend}};

\end{tikzpicture}

%% file: neurips_2020.bbl
\begin{thebibliography}{10}

\bibitem{Pope2000}
Stephen~B Pope.
\newblock {\em {Turbulent Flows}}.
\newblock Cambridge University Press, 2000.

\bibitem{Pope1985}
Stephen~B. Pope.
\newblock {PDF methods for turbulent reactive flows}.
\newblock {\em Prog. Energy Combust. Sci.}, 11:119--192, 1985.

\bibitem{Jenny2001b}
Patrick Jenny, Stephen~B. Pope, Metin Muradoglu, and David~A. Caughey.
\newblock {A Hybrid Algorithm for the Joint PDF Equation of Turbulent Reactive
  Flows}.
\newblock {\em Journal of Computational Physics}, 166(2):218--252, 2001.

\bibitem{Duraisamy2019}
Karthik Duraisamy, Gianluca Iaccarino, and Heng Xiao.
\newblock {Turbulence Modeling in the Age of Data}.
\newblock {\em Annual Review of Fluid Mechanics}, 51(1):357--377, 2019.

\bibitem{Tracey2013}
Brendan Tracey, Karthik Duraisamy, and Juan~J. Alonso.
\newblock {Application of supervised learning to quantify uncertainties in
  turbulence and combustion modeling}.
\newblock {\em 51st AIAA Aerospace Sciences Meeting including the New Horizons
  Forum and Aerospace Exposition 2013}, (January):1--18, 2013.

\bibitem{Ling2016}
Julia Ling, Andrew Kurzawski, and Jeremy Templeton.
\newblock {Reynolds averaged turbulence modelling using deep neural networks
  with embedded invariance}.
\newblock {\em Journal of Fluid Mechanics}, 807:155--166, 2016.

\bibitem{Beck2019}
Andrea Beck, David Flad, and Claus~Dieter Munz.
\newblock {Deep neural networks for data-driven LES closure models}.
\newblock {\em Journal of Computational Physics}, 398, 2019.

\bibitem{Novati2020}
Guido Novati, Hugues~Lascombes de~Laroussilhe, and Petros Koumoutsakos.
\newblock {Automating Turbulence Modeling by Multi-Agent Reinforcement
  Learning}.
\newblock {\em arXiv preprint arXiv:2005.09023.}, 2020.

\bibitem{Haworth1986}
Daniel~C. Haworth and Stephen~B. Pope.
\newblock {A generalized Langevin model for turbulent flows.}
\newblock {\em The Physics of Fluids}, 29(2):387--405, 1986.

\bibitem{Dinh2014}
Laurent Dinh, David Krueger, and Yoshua Bengio.
\newblock {NICE: Non-linear Independent Components Estimation}.
\newblock {\em arXiv preprint arXiv:1410.8516.}, 2014.

\bibitem{Rezende2015}
Danilo~Jimenez Rezende and Shakir Mohamed.
\newblock {Variational Inference with Normalizing Flows}.
\newblock {\em Proceedings of Machine Learning Research}, 37:1530--1538, 2015.

\bibitem{Dinh2017}
Laurent Dinh, Jascha Sohl-Dickstein, and Samy Bengio.
\newblock {Density estimation using Real NVP}.
\newblock {\em arXiv preprint arXiv:1605.08803}, 2016.

\bibitem{Ardizzone2019}
Lynton Ardizzone, Carsten L{\"{u}}th, Jakob Kruse, Carsten Rother, and Ullrich
  K{\"{o}}the.
\newblock {Guided Image Generation with Conditional Invertible Neural
  Networks}.
\newblock {\em arXiv preprint arXiv:1907.02392}, 2019.

\bibitem{Champagne1970}
F.~H. Champagne, V.~G. Harris, and S.~Corrsin.
\newblock {Experiments on nearly homogeneous turbulent shear flow}.
\newblock {\em J. Fluid Mech.}, 41:81--139, 1970.

\bibitem{Tavoularis1981}
Stavros Tavoularis and Stanley Corrsin.
\newblock {Experiments in nearly homogenous turbulent shear flow with a uniform
  mean temperature gradient. Part 1}.
\newblock {\em J. Fluid Mech.}, 104:311--347, 1981.

\bibitem{Rogallo1981}
Robert~S. Rogallo.
\newblock {Numerical Experiments in Homogeneous Turbulence}.
\newblock {\em NASA Tech. Memo. 81315}, 1981.

\bibitem{Rogers1987}
Michael~M. Rogers and Parviz Moin.
\newblock {The structure of the vorticity field in homogeneous turbulent
  flows}.
\newblock {\em Journal of Fluid Mechanics}, 176:33--66, 1987.

\bibitem{Pumir1996}
Alain Pumir.
\newblock {Turbulence in homogeneous shear flows}.
\newblock {\em Physics of Fluids}, 8(11):3112--3127, 1996.

\bibitem{Isaza2009}
Juan~C. Isaza and Lance~R. Collins.
\newblock {On the asymptotic behaviour of large-scale turbulence in homogeneous
  shear flow}.
\newblock {\em Journal of Fluid Mechanics}, 637:213--239, 2009.

\end{thebibliography}
